\documentclass[12pt]{iopart}
\usepackage{epsfig,amsfonts,amssymb}
\usepackage{cite,axodraw}
\newcommand{\newc}{\newcommand}
\newc{\inmath}[1] {\ifmmode#1\else$#1$\fi}
\newc{\definmath}[2] {\def#1{\ifmmode#2\else$#2$\fi}}
\newc{\gev}{\,GeV}
\newc{\mev}{\,MeV}
\newc{\ra}{\rightarrow}
\definmath{\rpv}{\mathrm{\not\!R_p}}
\definmath{\rp}{\mathrm{R_p}}
\newc{\real}{\mathcal{R}e}
\newc{\alsm}{{\displaystyle \sum_{\alpha=1,2}}}
\newc{\besm}{{\displaystyle \sum_{\beta=1,2}}}
\newc{\al}{\alpha}\def\Etmiss{\slashchar{E}_T}
\definmath{\lampp}{\lambda^{\prime \prime}}
\newc{\sgn}{\rm{sgn}\,}
\newc{\be}{\beta}
\newc{\ga}{\gamma}
\newc{\de}{\delta}
\newc{\sla}{\!\!\!\!\!\not\:\:\!}
\newc{\slab}{\!\!\!\!\!\not\,\,\,}
\newc{\slac}{\!\!\!\!\!\!\!\not\,\,\,\,}
\newc{\met}{$\not\!\!E_T$}
\newc{\cw}{\cos\theta_W}
\newc{\sw}{\sin\theta_W}
\newc{\ssw}{\sin^2\theta_W}
\newc{\ccw}{\cos^2\theta_W}
\newc{\cbe}{\cos\beta}
\newc{\sbe}{\sin\beta}
\newc{\ort}{\frac1{\sqrt{2}}}
\newc{\sh}{\hat{s}}
\def\slashchar#1{\setbox0=\hbox{$#1$}           
   \dimen0=\wd0                                 
   \setbox1=\hbox{/} \dimen1=\wd1               
   \ifdim\dimen0>\dimen1                        
      \rlap{\hbox to \dimen0{\hfil/\hfil}}      
      #1                                        
   \else                                        
      \rlap{\hbox to \dimen1{\hfil$#1$\hfil}}   
      /                                         
   \fi} 
\newc{\uh}{\hat{u}}
\newc{\tha}{\hat{t}}
\newc{\sa}{\sin\al}
\newc{\ca}{\cos\al}
\newc{\mz}{M_{\rm{Z}}}
\newc{\mw}{M_{\rm{W}}}
\definmath{\bv}{\mathrm{\not\!B}}
\definmath{\lv}{\mathrm{\not\!L}}
\newc{\beq}{\begin{equation}}
\newc{\eeq}{\end{equation}}
\newc{\ie}{{\it i.e.\/}\ }
\definmath{\lam}{\lambda}
\definmath{\cht}{\tilde{\chi}}
\definmath{\chgone}{\cht^+_1}
\definmath{\ntlone}{\cht^0_1}
\definmath{\ntltwo}{\cht^0_2}
\definmath{\sslr} {\tilde{\ell}_{R}}
\definmath{\glt}{\tilde{\rm{g}}}
\definmath{\upt}{\tilde{\rm{u}}}
\definmath{\qkt}{\tilde{\rm{q}}}
\definmath{\elt}{\tilde{\ell}}
\definmath{\hgt}{\tilde{\rm{H}}}
\definmath{\nut}{\tilde{\nu}}
\definmath{\dnt}{\tilde{d}}
\definmath{\ftl}{\rm{\tilde{\rm{f}}}}
\definmath{\psb}{\bar{\psi}}
\definmath{\rtt}{\sqrt{2}}
\definmath{\mut}{\tilde{\mu}}
\newc{\bath}{\bar{\theta}}
\newc{\tht}{\theta}
\newc{\JC}{{\bf J}}
\newc{\lra}{\longrightarrow}
\newc{\eg}{{\it e.g.\  }}
\newc{\barr}{\begin{eqnarray}}
\newc{\earr}{\end{eqnarray}}
\newc{\me}{\mathcal{M}}
\definmath{\dbm}{\partial_\mu}
\definmath{\dbmu}{\stackrel{\leftrightarrow\  }{\partial^\mu}}
\definmath{\sgm}{\sigma_\mu}\newc{\captionB}[2]{\caption[{#1}]{{\small {#2}}}}
\begin{document}
\title{Supersymmetric Models and Collider Signatures}
\author{I. Hinchliffe$^\dag$ and P. Richardson$^{\ddag,\S}$}
\address{$^\dag$\ Lawrence Berkeley National Laboratory, Berkeley, CA, USA}
\address{$^\ddag$\ Cavendish Laboratory, University of Cambridge, Madingley Road,
        Cambridge, UK}
\address{\S\ DAMTP, Centre for Mathematical Sciences, Wilberforce Road, Cambridge, UK}
\eads{\mailto{I\_Hinchliffe@lbl.gov}, \mailto{richardn@hep.phy.cam.ac.uk}}
\begin{abstract}

  We briefly review the SUGRA, GMSB and AMSB supersymmetry breaking models.
  We then discuss
  the phenomenological differences between them and consequent
  characteristic  experimental
  signatures. This is followed by a review of the discovery potential for
  supersymmetry at the Tevatron, LHC and a future $\rm{e}^+\rm{e}^-$ linear collider.

\end{abstract}

\section{Introduction}
\label{sect:intro}
  
  In the past twenty years there has been a great deal of theoretical 
study of low
  energy supersymmetry and a number of experimental searches. However,
  these
  searches have
  found no evidence for supersymmetry. Within the next ten years a
  number 
of new
  collider experiments will probe higher energies and low energy 
supersymmetry will
  either be discovered experimentally or will no longer be relevant to
  the problem of 
  electroweak symmetry breaking.

  After briefly surveying  the models which are used in experimental studies
  we will  discuss the discovery potential and signatures 
  for current and future experiments.

\section{SUSY Models}
\label{sect:SUSY}

  We will consider the Minimal Supersymmetric Standard Model (MSSM). 
This is the 
  supersymmetric extension of the the Standard Model (SM) which has minimal
  particle content, \ie it contains two Higgs doublets rather than the one of
  the SM and the superpartners of the Standard Model fields.
  This model has one less  parameter than  the Standard
  Model, 
provided
  that supersymmetry (SUSY) is unbroken.
  There is an additional parameter $\mu$ which gives mixing between
  the 
two Higgs doublets
  but the couplings in the scalar potential are constrained by SUSY.
  However, as the superpartners have not been
  observed,  SUSY must be broken in such a way as to not reintroduce the quadratic
  dependence of the Higgs mass on the cutoff scale. This is called soft SUSY breaking
  and leads to a large number of additional parameters:
  soft SUSY breaking masses for the
  scalars and gauginos; soft SUSY breaking
  $A$ terms, which couple two sfermions and a Higgs; and $B$ terms, which
  couple two Higgs bosons. The trilinear $A$ terms are
  only important for the third generation sfermions as they enter
  in terms proportional to the fermion masses. The Z boson mass is given in terms
  of the other  parameters of the model.

  In the MSSM R-parity related to baryon ($B$) and lepton number ($L$)
  by \mbox{$R_P=(-1)^{3B+2S+L}$}, is assumed to be conserved
  which means that SUSY particles can only be produced in pairs and the lightest
  SUSY particle (LSP) is stable. As the LSP is stable it must be
  neutral as charged stable matter is excluded by cosmological arguments.
  This implies  that all SUSY events have two LSPs in them
  leading  to missing transverse energy in SUSY events because the LSPs
  do not interact in the detector. R-parity need not be conserved and this leads
  to very different experimental signatures \cite{Dreiner:1997uz}.

  It is difficult  to use the MSSM
  for detailed experimental studies due to the large
  number of free parameters it contains. Furthermore large ranges of
  the parameters are excluded  and a real model will surely have far
  fewer fundamental parameters. Therefore models of SUSY breaking are
  used which are motivated by theoretical ideas. All have a common
  feature; SUSY is broken in some hidden sector and then transmitted to
  the MSSM fields. The models differ in how this transmission is
  accomplished:

\begin{description}

\item[SUGRA]  In supergravity models \cite{SUGRA} all the scalar masses ($M_0$), 
        the gaugino masses ($M_{1/2}$), the $A$ and $B$ parameters
        are assumed to be unified  at the grand unified (GUT)
        scale \mbox{($\sim 10^{15}\,\rm{GeV}$)}.
        As the model predicts $M_{\rm{Z}}$ in terms of the other parameters it
        is possible to use $\tan\beta=v_1/v_2$, the ratio of the vacuum expectation
        values (VEVs) for the two Higgs doublets, and the known value of $M_{\rm{Z}}$ to
        fix $B$ and $|\mu|$. This leaves five parameters $M_0$, $M_{1/2}$, $A$, 
        $\sgn\mu$, $\tan\beta$ which completely determine the mass spectrum and
        decay patterns of the particles. The gluino mass ($M_{\glt}$) is strongly
        correlated with $M_{1/2}$ 
        and the slepton mass with $M_0$. The LHC experiments have defined
        several SUGRA points which are often used in simulations
        \cite{atlastdr,Abdullin:1998pm}.

\item[GMSB]  The Gauge Mediated SUSY Breaking model \cite{GMSB} aims to
        solve the
        flavour changing neutral current problem which is generically
        present in SUGRA models by using gauge interactions rather
        than gravity to transmit the SUSY breaking. The messenger sector consists 
        of some particles, $X$, which have SM interactions and are aware of SUSY breaking.
        The simplest choice is to have the messenger particles in complete 
        $SU(5)$ {\bf 5} or {\bf 10} representations to preserve the 
        GUT symmetry.
        The fundamental SUSY breaking scale $F$ must be such that
        $\sqrt{F}\lesssim10^{10}\,\rm{GeV}$ or SUGRA breaking will dominate.
        The gaugino masses occur at one-loop while the squark and slepton masses
        occur at two-loop. The LSP is an almost massless gravitino so that
        the sparticles decay as in a SUGRA model followed by the decay of 
        the next-to-lightest SUSY particle (NLSP) to the gravitino.
        The NLSP need not be neutral and its lifetime is model dependent.
        This model has six parameters: $\Lambda=F/M$ the scale for the SUSY masses;
        $M>\Lambda$ the messenger mass scale; $N_5\geq1$ the number of
        ${\bf 5}+{\bf\bar{5}}$ messenger fields; the ratio of the
        vacuum expectation values $\tan\beta$; $\sgn\mu$ the sign of the
        $\mu$ parameter; and $C_{\rm{grav}}\geq1$ which controls the NLSP lifetime.
\item[AMSB]  The super-conformal anomaly \cite{AMSB} is always present 
and predicts sparticle
        masses in terms of $M_{3/2}$, the gravitino mass. 
        The simplest version of this model predicts tachyonic sleptons and therefore
        some other SUSY breaking mechanism must be present in order to
        get a realistic spectrum. One way to do this is to add a universal scalar mass
        (mAMSB) or new very heavy fields (DAMSB). The (mAMSB) model has four
        parameters: $M_0$ the universal scalar mass; $M_{3/2}$ the gravitino mass;
        $\tan\beta$ the ratio of Higgs VEVs; and $\sgn\mu$.
        The DAMSB model has five parameters: $M$ the mass of the new fields;
        $n$ the number of new fields; $M_{3/2}$; $\tan\beta$; and $\sgn\mu$.
        The AMSB model has one important feature in that the LSP is mainly wino
        like and
        almost degenerate with the lightest chargino, $\cht^+_1$, but this
        feature is lost in DAMSB.
\end{description}
All of these models are implemented in the ISAJET event generator \cite{ISAJET}.
\begin{figure}
\begin{center}
\includegraphics[width=0.57\textwidth,angle=-90]{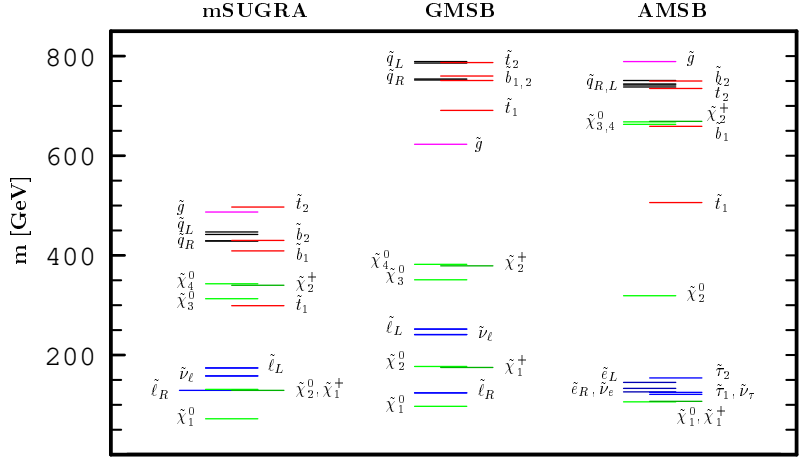}
\caption{Examples of mass spectra in SUGRA, GMSB and AMSB models for 
         $\tan\beta=3$ and $\sgn\mu=+$. The other parameters are: $M_0=100\,\rm{GeV}$,
        $M_{1/2}=200\,\rm{GeV}$ for the SUGRA model;
        $M=100\,\rm{TeV}$, $N_5=1$, $\Lambda=70\,\rm{TeV}$ for the GMSB model;
        $M_0=200\,\rm{GeV}$, $M_{3/2}=35\,\rm{TeV}$ for the AMSB model. Plot taken
        from \cite{tesla}.}
\label{fig:spectrum}
\end{center}
\end{figure}

  The spectra of the SUSY particles in these models can be very similar, 
  an example of each is shown in Fig.\,\ref{fig:spectrum}.
  The main differences in these models are the ratios of the squark to 
  slepton masses and the differences between the electroweak gaugino and gluino
  masses. 
  In the SUGRA model the scalar masses are universal at
  the GUT scale and therefore the differences in the masses come from
  the renormalization group evolution (RGE) to the electroweak scale.
  However, in the GMSB models the masses at the SUSY breaking scale are proportional
  to the relevant gauge couplings and therefore the strongly interacting squarks
  are heavier than the sleptons, relative to the SUGRA model. In the AMSB
  model the sleptons  are very light compared to the squarks unless
  the universal scalar mass is very large.
  Similarly for the gauginos the splitting of the gluino and electroweak gaugino
  masses in the SUGRA model comes from the RGE however in the GMSB model the 
  gaugino masses are proportional to the gauge couplings at the SUSY breaking scale and
  therefore the strongly interacting gluino is heavier than the weakly interacting
  gauginos. In the AMSB model the soft breaking mass for the gluino is
  larger than the soft breaking masses for the electroweak gauginos at the GUT scale
  which gives a bigger splitting between the gluino and electroweak gauginos masses
  than in SUGRA models. 
  The nature of the lightest neutralino is also different in the different models. 
  In the SUGRA model it is usually   mainly bino  whereas in the AMSB model
  it is mainly wino and degenerate with the lightest chargino.
  As discussed above, the  lightest neutralino in the GMSB model is
  the Gravitino and the NLSP, which can be neutral or charged, is most
  important for phenomenology.

  Obviously once supersymmetry is discovered it will be important to 
  make accurate measurements of SUSY particles masses and couplings
   in order to investigate the model of SUSY breaking. However
  at present, given we have seen no evidence for supersymmetry
  in experimental studies,  the main interest is in simulating
  models which give qualitatively different experimental signatures so
  that we can be certain that all variants can be observed.  For example
  the DAMSB model is very similar to a SUGRA model and has not
  therefore been subjected to many detailed studies,
 whereas in the mAMSB model which has an almost degenerate
  lightest neutralino and chargino the dominant chargino decay mode is
  $\cht^+_1\ra\pi^+\cht^0_1$ which is very different from SUGRA models
  and therefore of more interest.

  There are a number of signatures which are characteristic of supersymmetry,
  regardless of the model of SUSY breaking:
  missing transverse energy; a high multiplicity of high transverse momentum jets;
  many isolated leptons; copious b-jet production; a large rate of Higgs production;
  isolated photons; and quasi-stable charged particles. It should be noted that
  not all of these signals are present in all models and that production of 
  any heavy object will give some of these signals.

  In order to simulate supersymmetry it is essential to have a consistent model.
  We cannot consider one sparticle in isolation because all the supersymmetric particles
  which are kinematically allowed will be produced. 
  In hadron colliders production of the squarks and gluinos dominates provided the
  centre-of-mass energy is high enough. The production of those
  sparticles which only have electro-weak couplings may be dominated by the decays
  of squarks and not by direct production. The dominant backgrounds at the LHC
  are combinatorial from SUSY events after some simple cuts are
  applied.
 The situation at the Tevatron where
  electroweak gaugino production may dominate and Standard Model processes
  are the most important source of background is very different. 
  At a lepton  collider where the full spectrum is unlikely to 
  be accessible and the beam energy and polarization can be used to separate sparticles
  the situation   is   much  simpler.

\section{Tevatron}
\label{sect:tevatron}

\begin{figure}
\begin{center}
\includegraphics[width=0.78\textwidth]{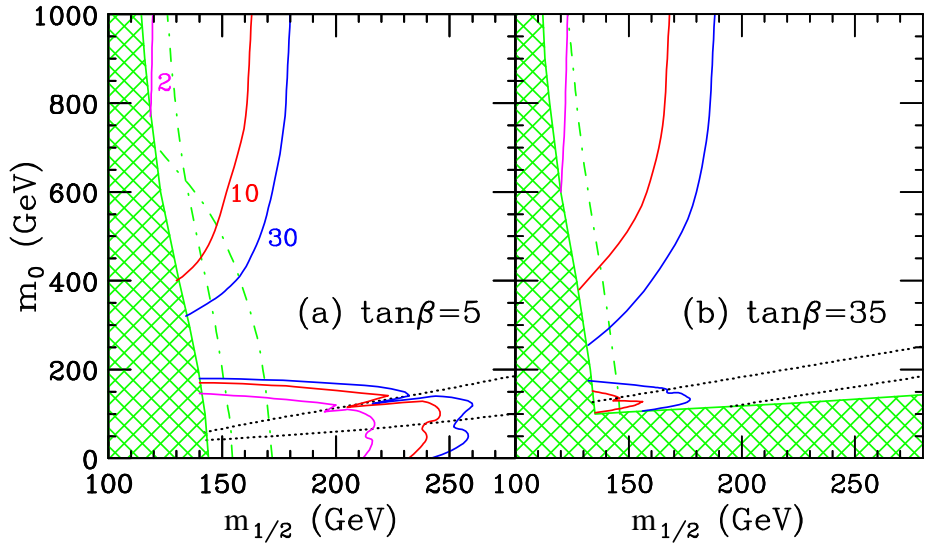}
\caption{Tevatron reach in the Tri-lepton channel in the $M_0,M_{1/2}$ plane, for fixed values
        of $A_0=0$, $\mu>0$ and {\bf (a)} $\tan\beta=5$ or {\bf (b)} $\tan\beta=35$.
        Results are shown for 2, 10 and 30~$\rm{fb}^{-1}$ integrated luminosity.
        The curves require the observation of at least 5 events and are $3\sigma$
        exclusion contours. The cross-hatched region is excluded by current limits
        on the superpartner masses and the dot-dashed lines correspond to the projected LEP-II
        reach for the chargino and lightest Higgs masses.
        Figure taken from \cite{Abel:2000vs,Matchev:1999yn}.}
\label{fig:tevatron1}
\end{center}
\end{figure}

\begin{figure}
\begin{center}
\includegraphics[width=0.8\textwidth]{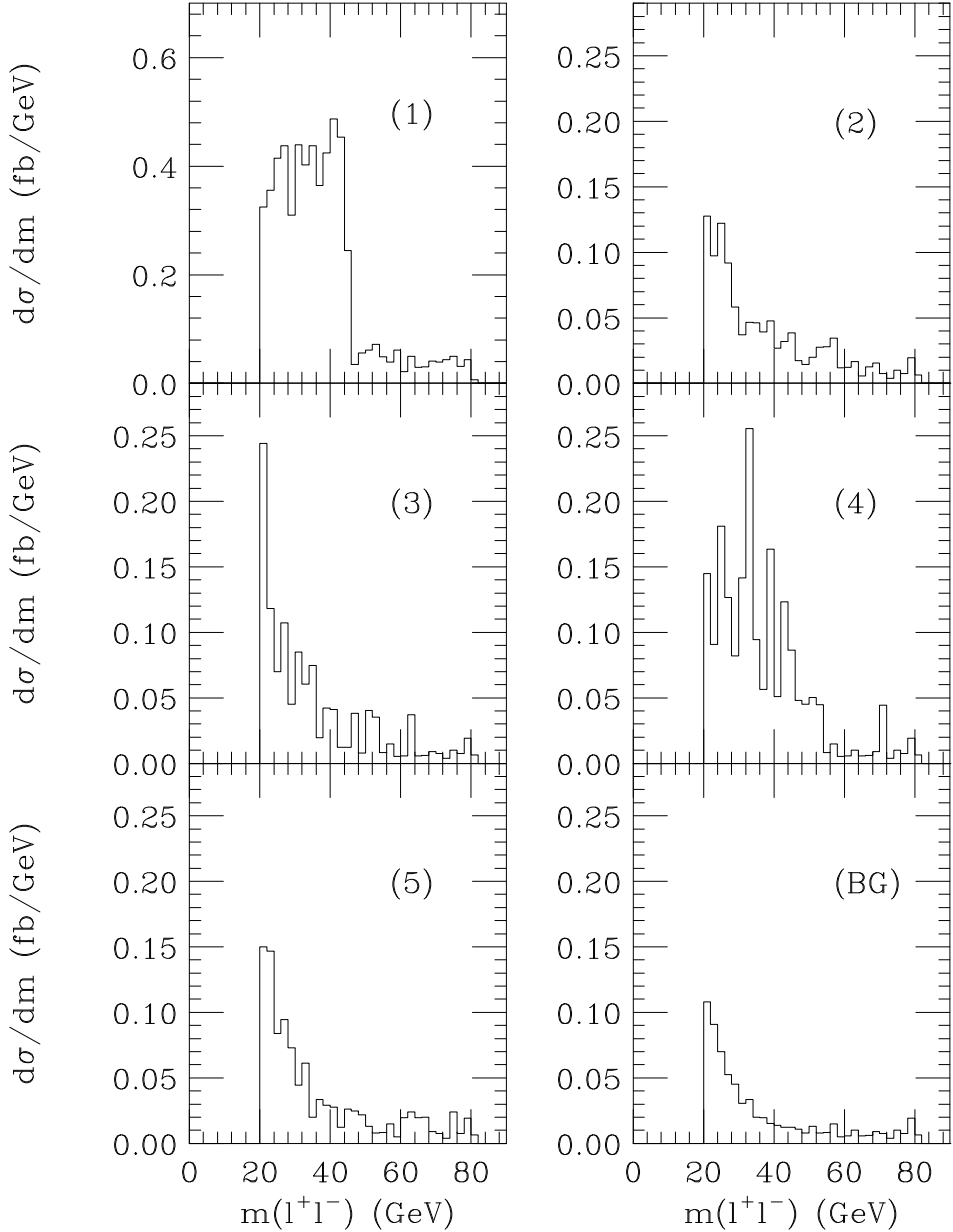}
\caption{Opposite sign, same flavour dilepton mass reconstruction for
        five study points given below and the WZ and $\rm{t}\rm{\bar{t}}$
        background. The background is included in all the histograms. It should
        be noted that since this plot was produced point 1 has been ruled out
        by searches at LEP.
        Point 1 has $M_0=100\,\rm{GeV}$, $M_{1/2}=200\,\rm{GeV}$, $A_0=0\,\rm{GeV}$, 
         $\tan\beta=3$, $\sgn\mu=+$;
        point 2 has $M_0=140\,\rm{GeV}$, $M_{1/2}=175\,\rm{GeV}$, $A_0=0\,\rm{GeV}$, 
         $\tan\beta=35$, $\sgn\mu=+$;
        point 3 has $M_0=200\,\rm{GeV}$, $M_{1/2}=140\,\rm{GeV}$, $A_0=-500\,\rm{GeV}$, 
         $\tan\beta=35$, $\sgn\mu=+$;
        point 4 has $M_0=250\,\rm{GeV}$, $M_{1/2}=150\,\rm{GeV}$, $A_0=-600\,\rm{GeV}$, 
         $\tan\beta=3$, $\sgn\mu=+$;
        point 5 has $M_0=150\,\rm{GeV}$, $M_{1/2}=300\,\rm{GeV}$, $A_0=0\,\rm{GeV}$, 
         $\tan\beta=30$, $\sgn\mu=-$ and non-universal GUT-scale Higgs masses 
        $M_{H_1,H_2}=500\,\rm{GeV}$. Plot taken from \cite{Abel:2000vs,Baer:1999bq}.}
\label{fig:tevatron2}
\end{center}
\end{figure}
  In Run I of the Tevatron neither CDF or D0 claimed the discovery of any signal for 
  supersymmetry. A number of limits were obtained which 
  we will not discuss here~\cite{Abe:1997yy,Abbott:1999xc}. 
  Run II of the Tevatron which has both an increase in the centre-of-mass energy to 
  2~TeV and one hundred times the luminosity will extend the mass range but the
  search reach is limited. Due to the centre-of-mass energy the production of squarks
  and gluinos may not dominate and the best channel for the
  discovery of SUSY may be  gaugino production, \mbox{\ie~production} of
  $\cht^0_2\cht^+_1\ra\ell^+\ell^-\cht^0_1\ell^+\nu\cht^0_1$.
  Fig.\,\ref{fig:tevatron1} shows the search potential in this channel at Run II of the
  Tevatron for different integrated luminosities. The background to this process
  is dominated by gauge boson pair production followed by leptonic
  decays with $\rm{W}\rm{Z}/\gamma$ being the dominant background process.

  If this signal is observed structure in the $\ell^+\ell^-$ mass distribution
  will constrain the $\cht^0_1$ and $\cht^0_2$ masses, this will be discussed in more
  detail for the LHC in the next section. However as can be seen from
  Fig.\,\ref{fig:tevatron2} for those points which are still allowed by the LEP
  limits the  endpoint of the distribution which gives information on the mass
  difference of the lightest two neutralinos will be difficult to measure.

  It is possible to extend the search reach by using channels
  involving jets and missing
  transverse energy.
  If supersymmetry is discovered at the Tevatron it will determine the mass scale of
  some of the particles. The LHC will then make detailed measurements
  of the SUSY spectrum.

\section{LHC}
\label{sect:LHC}

\begin{figure}
\begin{center}
\vspace{2cm}
\includegraphics[width=0.5\textwidth]{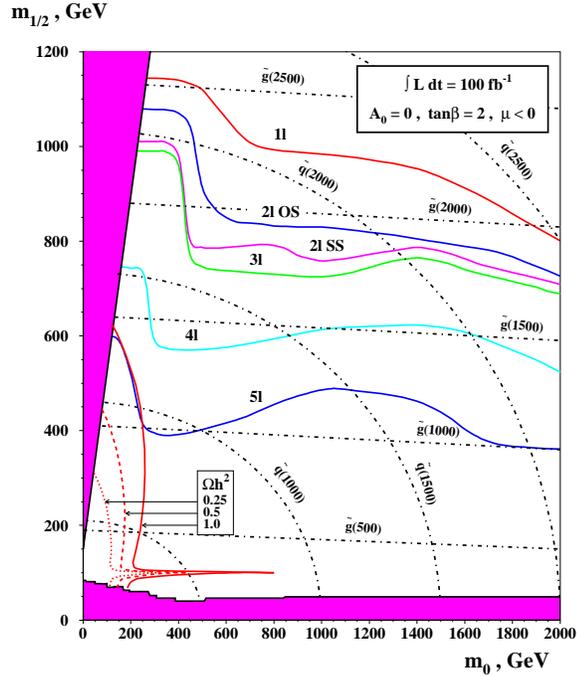}
\caption{Discovery potential of the LHC for final states with at least two jets,
         $\Etmiss >100\,\rm{GeV}$ and at least one isolated lepton.
         An integrated luminosity of $100\,\rm{fb}^{-1}$ has been assumed.
        Plot taken from \cite{Kharchilava:1997uw}.}
\label{fig:LHC1}
\end{center}
\end{figure}

  Many detailed studies of the search potential of the LHC have been performed by
  both the ATLAS \cite{atlastdr}
  and CMS \cite{Abdullin:1998pm} collaborations. At the LHC the strongly interacting 
  squarks and gluinos are predominantly produced unless they are very
  heavy
 giving large SUSY production 
  cross sections. This allows hard cuts to be applied to the data in order
  to reduce the Standard Model background leaving combinatorial backgrounds from
  the SUSY events themselves as the dominant source of background.
  Most of the studies have assumed the full integrated luminosity of the LHC
  which enables hard cuts to be used. In practice the discovery of SUSY will
  probably take less time and use weaker cuts in which case the Standard Model
  background will be more important.

  The LHC studies can be broadly grouped into two categories: the first attempts
  to find some signal indicative of SUSY and find an excess in order to discover
  supersymmetry by exploiting inclusive signatures; 
the second then tries to make use of more information
  from the SUSY events in order to measures masses and other parameters of the model.

   There is a very large range of accessible masses in inclusive signals, \ie jets,
  leptons and missing transverse energy ($\Etmiss$). 
  Fig.\,\ref{fig:LHC1}
  shows the mass reach in SUGRA models for the 
  CMS experiment~\cite{Kharchilava:1997uw}. This
  covers all the interesting theoretical range, 
  \mbox{\ie~$M_{\glt}\lesssim2.5\,\rm{TeV}$}.
  It is useful to define global variables for SUSY searches. For example if we consider
  events with at least four jets and missing transverse energy the variable
\beq
M_{\rm{eff}} = p_{T,1}+p_{T,2}+p_{T,3}+p_{T,4}+\Etmiss ,
\eeq
  where $p_{T,i}$ is the transverse momentum of the $i$th jet, is very useful.
  The peak in the
  $M_{\rm{eff}}$ distribution correlates well with the SUSY mass scale where
  $M_{\rm{SUSY}}=\min(M_{\upt},M_{\glt})$. This can determine the squark/gluino
  masses to about 15\% \cite{Tovey:2000wk}.

\begin{figure}
\begin{center}
\includegraphics[width=0.43\textwidth]{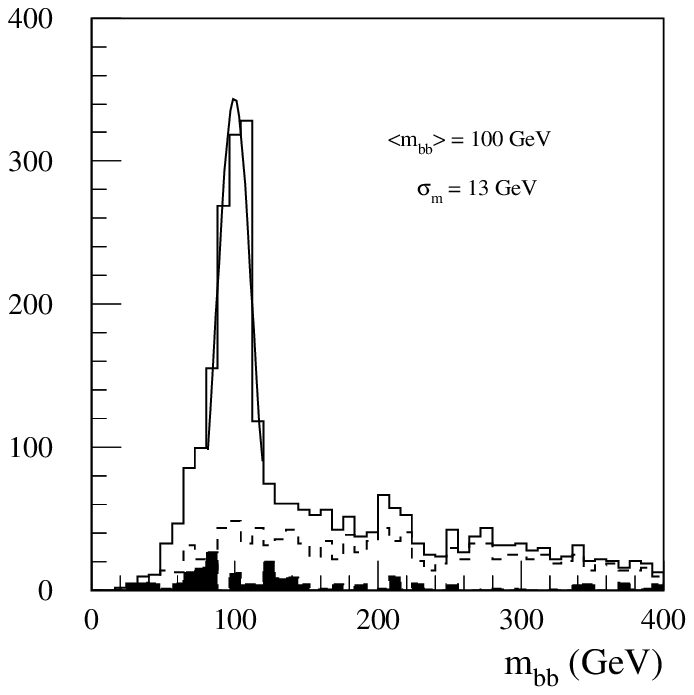}\hfill
\includegraphics[width=0.463\textwidth]{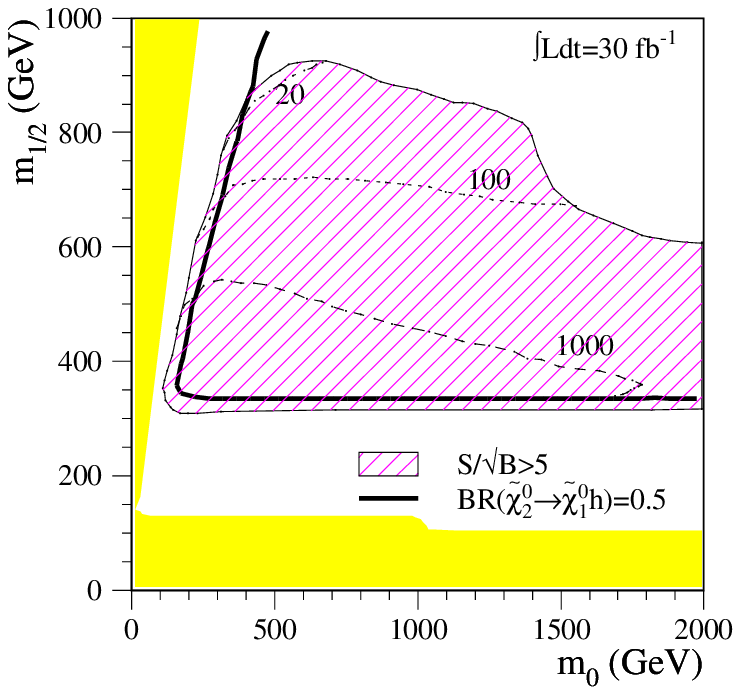}\\
\begin{picture}(0,0)
\Text(-185,190)[]{{\bf (a)}}
\Text(65,190)[]{{\bf (b)}}
\end{picture}
\caption{{\bf (a)}
        The reconstructed $\rm{b}\rm{\bar{b}}$ mass distribution for events
        passing $\rm{h}\ra\rm{b}\rm{\bar{b}}$ selection cuts. The distributions are
        for the SM background (shaded), the total SUSY+SM background (dashed) and the
        summed signal+background for SUGRA point 1.
        {\bf (b)} 
        The $5\sigma$ discovery contour curves for $\rm{h}\ra\rm{b}\rm{\bar{b}}$ from SUSY cascade decays in the $M_0,M_{1/2}$
        plane for $\tan\beta=10$, $\sgn\mu=+$. The expected number of reconstructed
        $\rm{h}\ra\rm{b}\rm{\bar{b}}$ events are also shown. The dark shaded 
        regions are theoretical or experimentally excluded.
        Both (a) and (b) assume $30\,\rm{fb}^{-1}$ integrated luminosity 
        and are taken from \cite{atlastdr}.}
\label{fig:LHCbb}
\end{center}
\end{figure}

   There have been many studies of techniques to reconstruct sparticle masses and
   properties. Here we will illustrate the techniques by choosing
   examples from 
case studies. In general both the squarks and gluino are produced. Then depending on
   the relative masses one decays into the other as these decays occur via
   the strong interaction.
   The weak gauginos are then produced in the decays of the lighter strongly interacting
   particles, for example $\qkt_L\ra\cht^0_2\rm{q}$.
   In most models a significant number of second-to-lightest neutralinos are 
   produced. These neutralinos then either decay via $\cht^0_2\ra\cht^0_1\rm{h}$
   or $\cht^0_2\ra\cht^0_1\ell^+\ell^-$, possibly via either an intermediate slepton
   $\cht^0_2\ra\elt^+\ell^-\ra\cht^0_1\ell^+\ell^-$ or Z boson 
   $\cht^0_2\ra\rm{Z}^0\cht^0_1\ra\cht^0_1\ell^+\ell^-$. The Higgs decay mode
   tends to dominate if it is kinematically accessible.
   Most studies have used these decays as a starting point for mass measurements.
   Many other SUSY particles can  then be identified by adding more jets or leptons
   to reconstruct other particles in the decay chain.

   If the decay of the $\cht^0_2\ra\cht^0_1\rm{h}$ exists then approximately $20\,\%$ of
   SUSY events contain $\rm{h}\ra\rm{b}\rm{\bar{b}}$. In these models the Higgs
   would be discovered in SUSY events at the LHC rather than by the traditional
   Standard Model-like Higgs searches. 
   The mass of $\rm{b}\bar{\rm{b}}$ pairs is shown in
   Fig.\,\ref{fig:LHCbb}(a)
 after the following
   cuts: $\Etmiss >300\,\rm{GeV}$; more than two jets
   with $p_T>100\,\rm{GeV}$ and more than one with $|\eta|<2$; no isolated leptons
   to suppress the $\rm{t}\rm{\bar{t}}$ background; only two b-jets with
   $p_{T,\rm{b}}>55\,\rm{GeV}$ and $|\eta|<2$; $\Delta R_{\rm{b}\rm{\bar{b}}}<1.0$
   again to suppress the $\rm{t}\rm{\bar{t}}$ background. This gives a clear
   peak in the $\rm{b}\rm{\bar{b}}$ mass distribution at the  Higgs mass. The SM
   background is very small and the dominant background is from other SUSY decays.
   This method works over a large region of parameter space in SUGRA Models, 
   Fig.\,\ref{fig:LHCbb}(b).

\begin{figure}
\begin{center}
\includegraphics[width=0.43\textwidth]{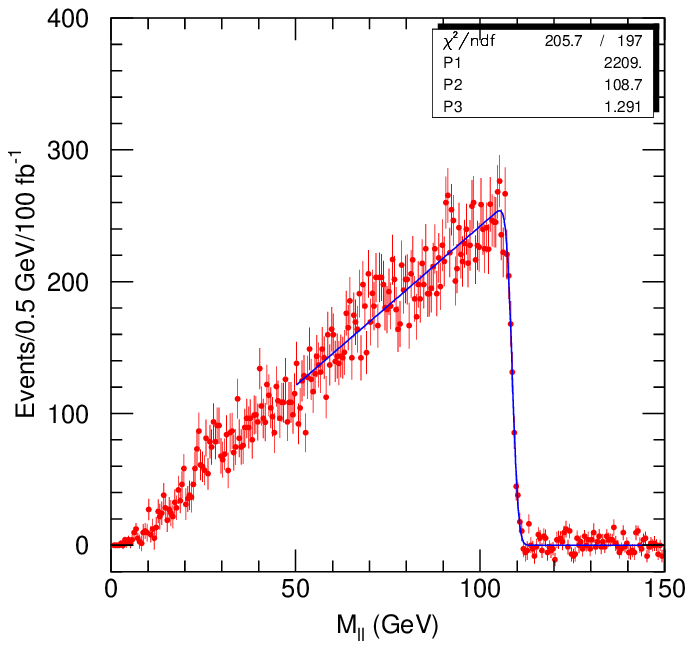}\hfill
\includegraphics[width=0.43\textwidth]{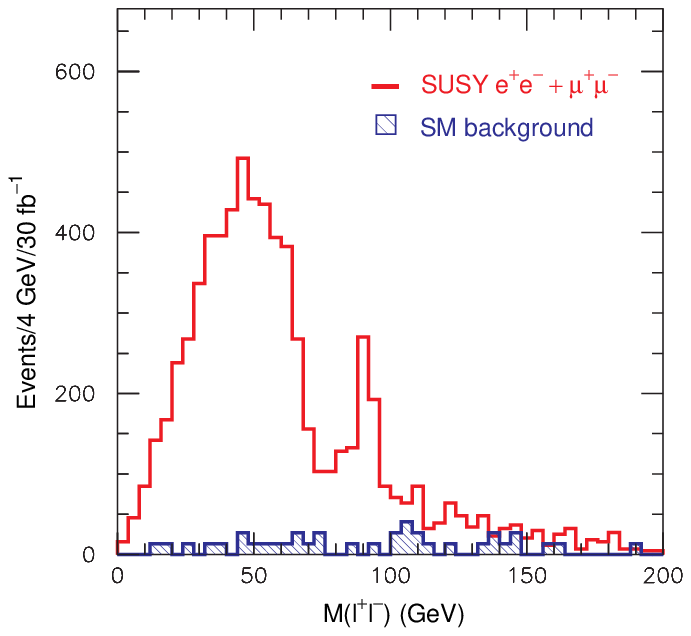}\\
\begin{picture}(0,0)
\Text(-175,175)[]{{\bf (a)}}
\Text(82,175)[]{{\bf (b)}}
\end{picture}
\vspace{-5mm}
\caption{{\bf (a)} Dilepton $\cht^0_2\ra\elt^\pm_R\ell^\mp\ra\cht^0_1\ell^+\ell^-$ signal at
        SUGRA point 5 for $100\,\rm{fb}^{-1}$ 
        including backgrounds after flavour subtraction, \ie the plot shows
        \mbox{$\rm{e}^+\rm{e}^-+\mu^+\mu^--e^\pm\mu^\mp$}.
        {\bf (b)} Dilepton $\cht^0_2\ra\cht^0_1\ell^+\ell^-$ distribution for 
         SUGRA point 4 (solid) and SM background (shaded) with $30\,\rm{fb}^{-1}$.
        Both plots are taken from \cite{atlastdr}. }
\label{fig:mll}
\end{center}
\end{figure}

  In the regions of parameter space where the decay mode $\cht^0_2\ra\cht^0_1\rm{h}$
  is not kinematically accessible, the reconstruction of the leptonic decay mode is
  necessary in order to constrain the $\cht^0_2$ mass.
  The important decay modes are either via a real slepton, Fig.\,\ref{fig:mll}(a),
  or via virtual sleptons and gauge bosons Fig.\,\ref{fig:mll}(b). 
  In Fig.\,\ref{fig:mll}b there is also a peak at the Z mass due to the
  production of Z bosons in other SUSY decays.

\begin{figure}
\begin{center}
\includegraphics[width=0.3\textwidth]{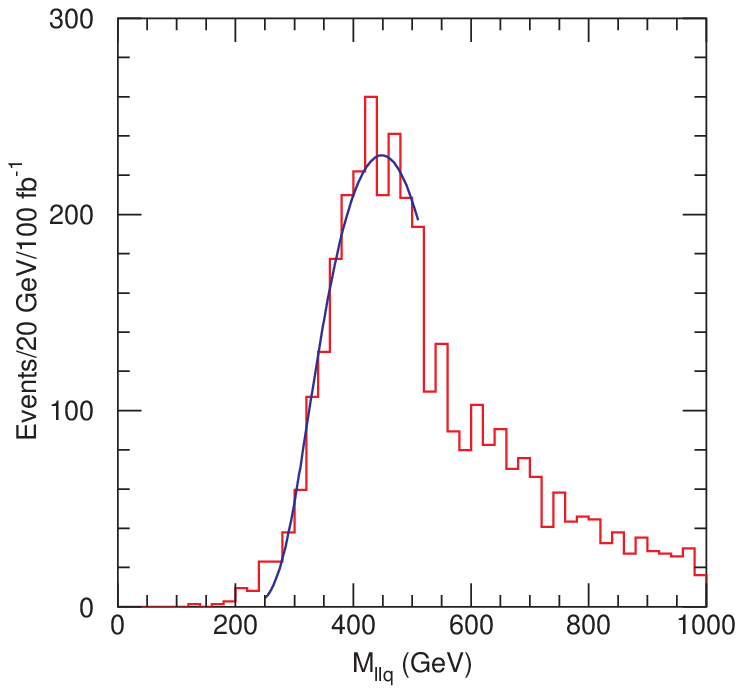}\hfill
\includegraphics[width=0.309\textwidth]{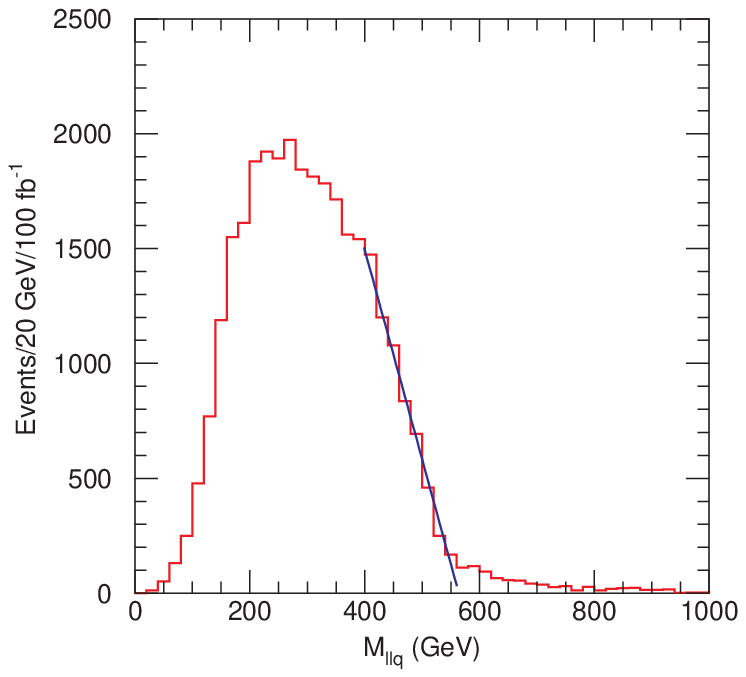}\hfill
\includegraphics[width=0.3\textwidth]{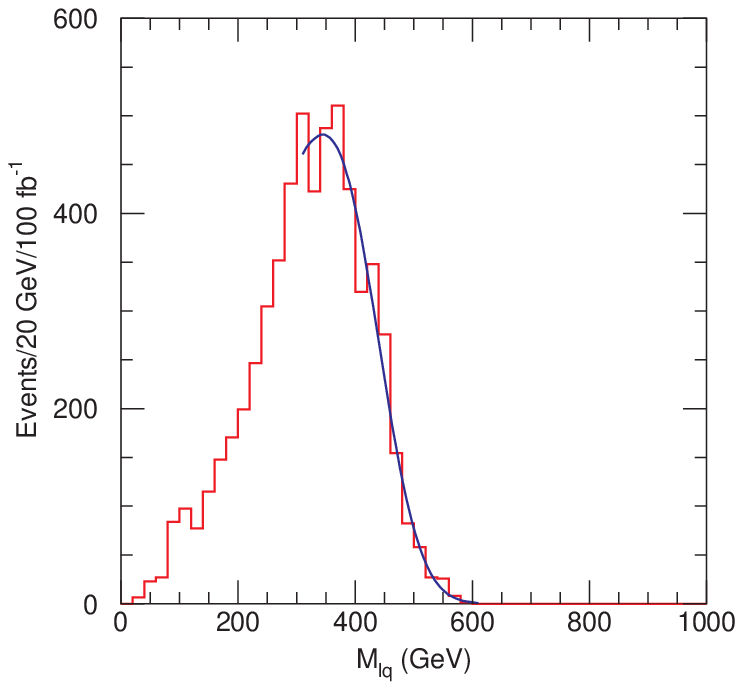}\\
\begin{picture}(0,0)
\Text(-183,125)[]{{\bf (a)}}
\Text(-28,125)[]{{\bf (b)}}
\Text(125,125)[]{{\bf (c)}}
\end{picture}
\vspace{-5mm}
\caption{{\bf (a)} Distribution of the larger $\ell^+\ell^-\rm{q}$ mass 
        for $M_{\ell\ell}>M^{\rm{max}}_{\ell\ell}/\sqrt{2}$.
        {\bf (b)} The smaller of the two $\ell^+\ell^-\rm{q}$ masses for the
        signal.
        {\bf (c)} The $\ell^\pm\rm{q}$ mass distribution for combinations with
         $M_{\ell^+\ell^-\rm{q}}<600\,\rm{GeV}$.
         Plots taken from \cite{atlastdr}.}
\label{fig:llqedges}
\end{center}
\end{figure}

\begin{figure}
\begin{center}
\includegraphics[width=0.45\textwidth]{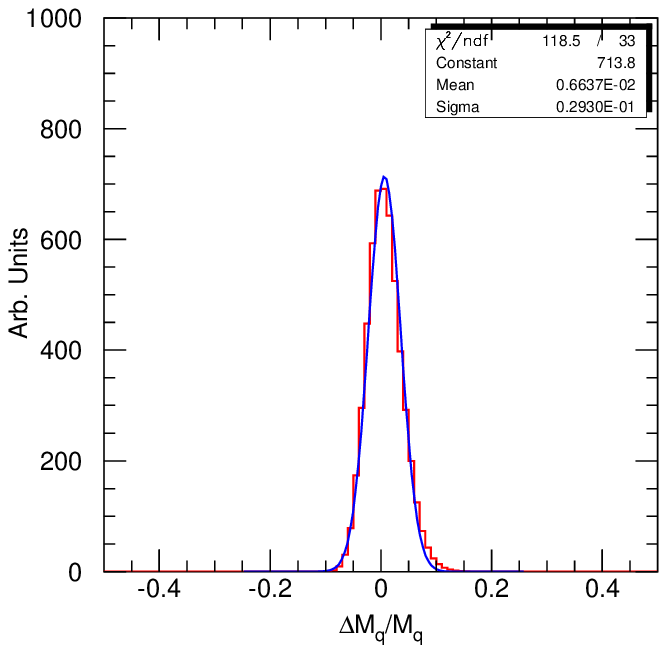}\hfill
\includegraphics[width=0.455\textwidth]{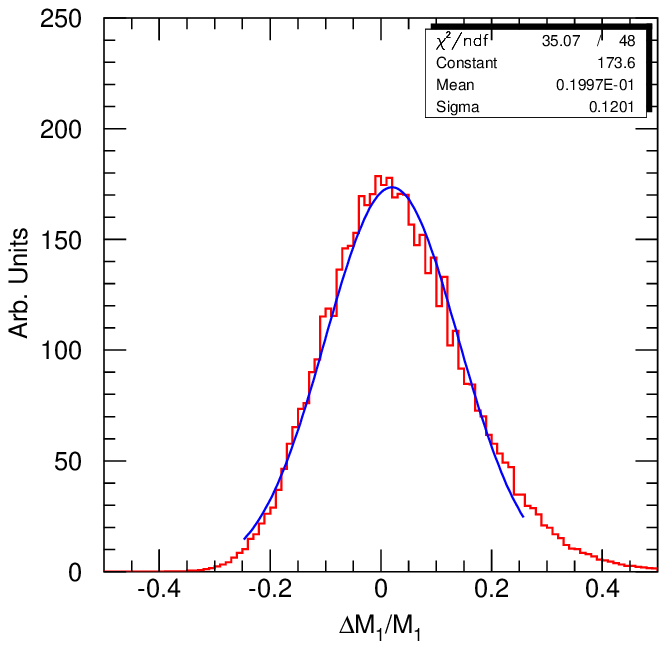}\\
\begin{picture}(0,0)
\Text(-170,185)[]{{\bf (a)}}
\Text(75,185)[]{{\bf (b)}}
\end{picture}
\vspace{-5mm}
\caption{{\bf (a)} Distribution of the fractional difference between
         the reconstructed and true squark mass.
        {\bf (b)} Distribution of the fractional difference between the reconstructed
         and true $\cht^0_1$ mass. Both plots were taken from \cite{atlastdr}.}
\label{fig:masses}
\end{center}
\end{figure}

  The leptons produced in neutralino decays can  be combined with the other
  decay products of the heavier SUSY particles in order to reconstruct them.
  For example the $\cht^0_2$ is often produced in left squark decays, 
  $\qkt_L\ra \rm{q}\cht^0_2\ra\rm{q}\elt\ell\ra\rm{q}\ell^+\ell^-\cht^0_1$.
  In order to identify this decay chain the following cuts are applied:
  two isolated leptons with transverse momentum, $p_T>10\,\rm{GeV}$;
  more than four jets one with $p_T>100\,\rm{GeV}$ and the rest with
  $p_T>50\,\rm{GeV}$; missing transverse energy
  $\Etmiss >\max(100.0,0.2M_{\rm{eff}})$. The mass of the
  $\rm{q}\ell\ell$ system has a maximum at
\beq
 M^{\rm{max}}_{\ell\ell\rm{q}} = 
        \left[\frac{\left(M^2_{\qkt_L}-M^2_{\cht^0_2}\right)
                    \left(M^2_{\cht^2_0}-M^2_{\cht^0_1}\right)}
        {M^2_{\cht^0_2}}\right]^{\frac12}\!\!,
\eeq
  where $M^2_{\qkt_L}$ is the squark mass and $M_{\cht^0_{1,2}}$ are the
  lightest and next-to-lightest neutralino masses respectively. This 
  distribution is shown in Fig.\,\ref{fig:llqedges}(a).
  The minimum of this distribution, Fig.\,\ref{fig:llqedges}(b),
  and the $\ell\rm{q}$ distribution, Fig.\,\ref{fig:llqedges}(c),
  also provide useful information. This system has four constraints, \ie
  the upper edges in $\ell\rm{q}$, $\ell\ell$ and 
  $\ell\ell\rm{q}$ distributions, 
  and the lower edge in the $\ell\ell\rm{q}$ distribution, and
  four unknowns, \ie $M_{\cht^0_1}$, $M_{\cht^0_2}$, $M^2_{\qkt_L}$
  and $M^2_{\elt_R}$ which can therefore be reconstructed.
  The errors on the $\cht^0_1$, $\cht^0_2$, $\qkt_L$ and 
  $\elt_R$ masses are 12\,\%, 6\,\%,
  3\,\% and 9\,\%, respectively. The mass of the unobserved LSP
  is determined, Fig.\,\ref{fig:masses}.
   The errors on the particle masses are strongly correlated and
  a precise determination of one mass would reduce the errors on the rest
  \cite{Allanach:2000kt}.
  
  In GMSB models a similar type of analysis is possible. For example,
  consider a case where
  the NLSP is the lightest neutralino which then decays to a gravitino and
  a photon, \ie $\cht^0_1\ra\gamma\tilde{\rm{G}}$. As in SUGRA models 
  the $\cht^0_2$ can decay leptonically to give the lightest neutralino. This
  gives the decay chain, $\cht^0_2\ra\ell^+\ell^-\cht^0_1\ra\ell^+\ell^-\gamma\tilde{\rm{G}}$.
  If we require $M_{\rm{eff}}>400\,\rm{GeV}$;
  $\Etmiss >0.1M_{\rm{eff}}$; at least two leptons and two photons
  where photons and electrons have $p_T>20\,\rm{GeV}$ and muons $p_T>5\,\rm{GeV}$.
  Here information can be obtained from the $\ell^+\ell^-$, $\ell^+\ell^-\gamma$
  and $\ell^\pm\gamma$ mass distributions. There are two structures in the
  $\ell^\pm\gamma$ distribution and therefore these distributions provide
  four constraints which  
  are sufficient to measure the $\cht^0_1$, $\cht^0_2$ and $\elt_R$ masses in this
  model.

  In AMSB models the signatures are very similar to SUGRA models except that the
  $\cht^+_1$ may not be observable due to the small mass difference between
  the lightest chargino and the LSP.

  The signatures of SUSY in models with R-parity violation are very different
  because the LSP can decay inside the detector which means there may be no missing
  transverse energy. The LSP will decay to give either leptons 
  \cite{atlastdr,Mirea:1999fs}, leptons  and jets \cite{atlastdr}
  or just jets \cite{atlastdr,Allanach:2001xz}. In these models the LSP
  mass can be found by reconstructing all its decay products which
  often allows the LSP mass to be measured with greater precision than is possible
  in SUGRA models.

\section{Lepton Colliders}
\label{sect:lepton}
  Any future linear collider will start operation after the LHC and therefore
  the study of supersymmetry at such a machine will be in the context of what we
  already know from the LHC. A collider with centre-of-mass energy of less
  than 1\,TeV will probably concentrate on the sparticles which only interact weakly,
  \ie the electroweak gauginos and sleptons.
  A higher energy collider  may also be able to produce
  the squarks, however at any lepton collider production of gluinos is difficult,
  unless they are lighter than the squarks in which case they are produced in
  squark decays.

  An $\rm{e}^+\rm{e}^-$ machine has a number of advantages over a hadron collider:
  the energy of the beam provides a kinematic 
  constraint; polarization of the electrons  and possibly positrons 
 provides a powerful tool and the signal to background ratio
  is of order one before cuts are applied. The great power of such a
  facility is the ability to make precise measurements of the masses
  and couplings of sparticles which may already have been observed at the LHC.
   All the examples we will consider are taken from
  \cite{tesla}. Similar studies can also be found in \cite{orange}.

\begin{figure}
\begin{center}
\includegraphics[width=0.5\textwidth,angle=90]{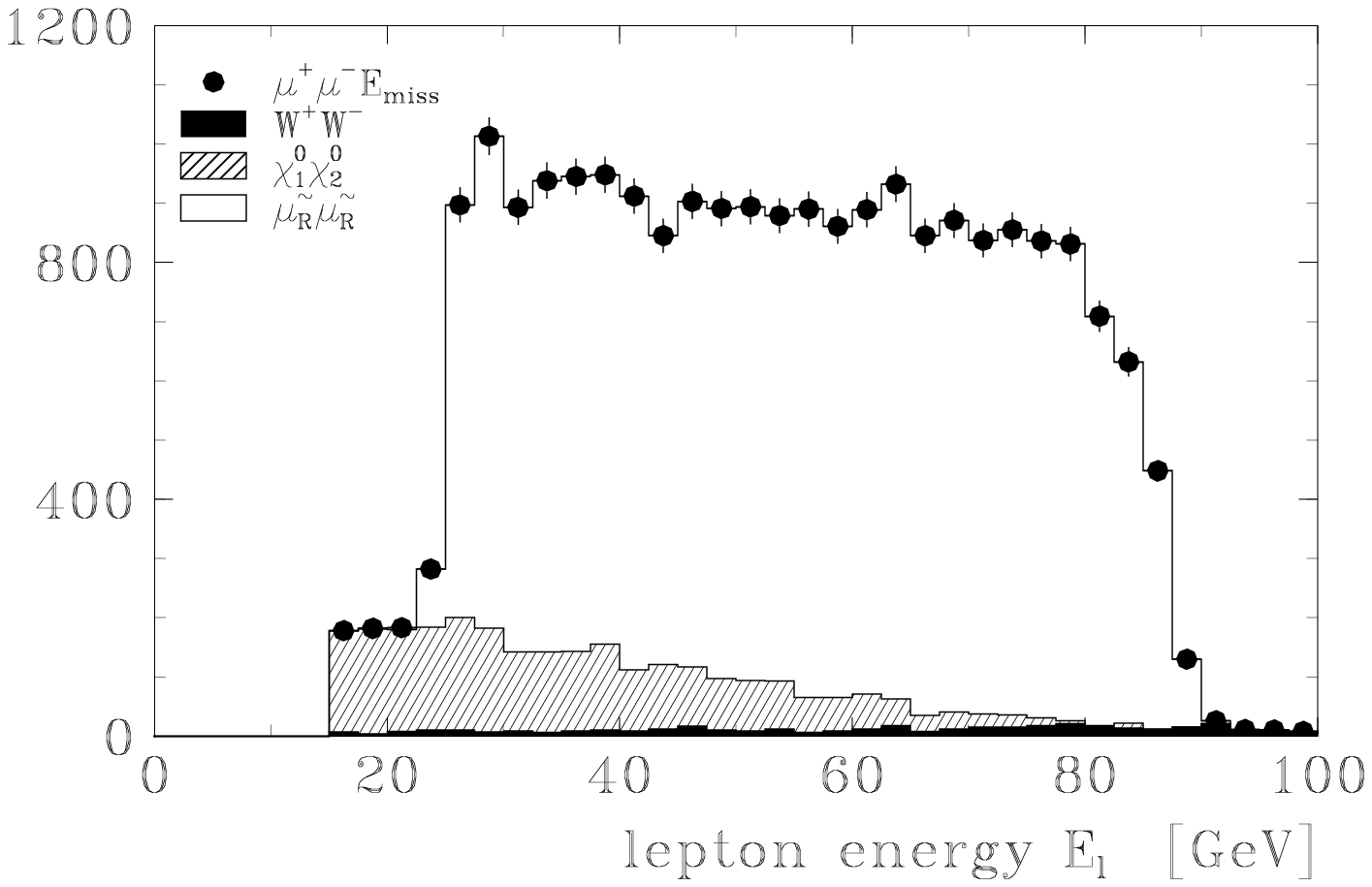}
\caption{The energy spectrum $E_\mu$ of the muons produced in the process
        \mbox{$\rm{e}^+\rm{e}^-\ra\mut_R\mut_R\ra\mu^-\cht^0_1\mu^+\cht^0_1$}
        at $\sqrt{s}=320\,\rm{GeV}$ for an integrated luminosity of
        $160\,\rm{fb}^{-1}$. Plot taken from \cite{tesla}.}
\label{fig:lepton1}
\end{center}
\end{figure}

  If we consider, for example, the production of smuons, 
  $\rm{e}^+\rm{e}^-\ra\mut^+\mut^-\ra\cht^0_1\mu^+\cht^0_1\mu^-$,
  the events will have a $\mu^+\mu^-$ pair and missing energy.
  The energy spectrum of the muons is shown Fig.\,\ref{fig:lepton1}. This spectrum 
  is flat apart from beamstrahlung, initial-state radiation and resolution effects
  at the high edge. The end points of this distribution can be related to the masses
  of the smuon and lightest neutralino.  Using this process both the $\cht^0_1$ and $\mut$ masses can 
  be determined to $\sim0.5\,\%$. If the polarization of the beams is changed
  the amount of left and right slepton in the mass eigenstate can also be determined.
In the case shown in Fig.\,\ref{fig:lepton1} the machine is below threshold for the
  other SUSY particles, apart from $\cht^0_1\cht^0_2$,
  and therefore the SUSY background is small. The Standard Model background
  is even smaller after event selection.
 
\begin{figure}
\begin{center}
\includegraphics[width=0.6\textwidth,trim = 110 20 0 120,clip,angle=-90]{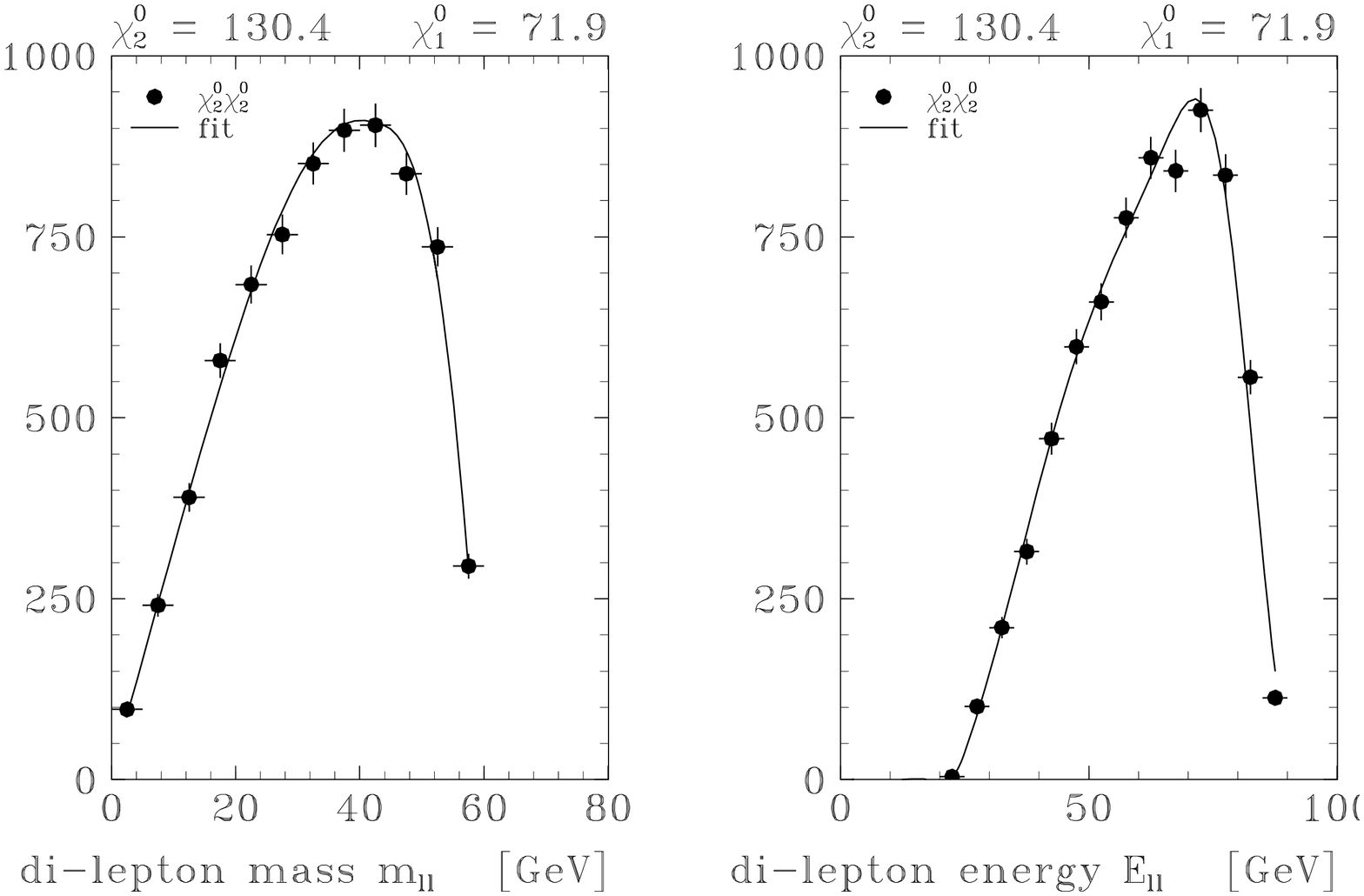}\\
\begin{picture}(0,0)
\Text(-38,235)[]{{\bf (a)}}
\Text(170,235)[]{{\bf (b)}}
\end{picture}
\vspace{-5mm}
\caption{{\bf (a)} The di-lepton mass and {\bf (b)} di-lepton energy spectrum
        for 
        \mbox{$\rm{e}^+\rm{e^-}\ra\cht^0_2\cht^0_2\ra\ell^+\ell^-\cht^0_1\ell^+\ell^-\cht^0_1$}
         at $\sqrt{s}=320\,\rm{GeV}$ for an integrated luminosity of $160\,\rm{fb}^{-1}$. Plot taken
        from \cite{tesla}. }
\label{fig:lepton2}
\end{center}
\end{figure}

  The heavier gauginos can also be produced, 
  \eg
  $\rm{e}^+\rm{e}^-\ra\cht^0_2\cht^0_2\ra\ell^+\ell^-\ell^+\ell^-\cht^0_1\cht^0_1$ at sufficiently high centre-of-mass energy.
  The dilepton masses and energies for this process are shown in
  Fig.\,\ref{fig:lepton2}. The masses of both the $\cht^0_1$ and $\cht^0_2$ can be
  determined from the di-lepton energy spectrum with typical errors of about $0.2\%$.
  The di-lepton mass spectrum gives additional information on the mass difference 
  between the two leptons and this difference can typical be measured with a precision
  of better than $50\,\rm{MeV}$.

  These mass measurements will be more accurate than those achievable by the LHC,
  if the lepton collider has sufficient energy to produced the
  particles. Furthermore it should be possible to determine the
  mixings in  the gaugino sector which is difficult at the LHC. Note
  that, as discussed above, the extraction of masses at the LHC may
  result in strongly correlated errors. Measurements from a linear
  collider of some masses could therefore be used to 
improve the errors on the LHC results for the
  heavier particles, \eg the squarks, which the lepton collider may not be able to produce.

\section{Conclusions}

  The upgraded Tevatron has some search potential for supersymmetry
  however, given the current limits, the event
  rates will be low. The most promising signal is from tri-leptons produced following 
  the production of $\cht^\pm_1\cht^0_2$, but the search range is limited.

  The production rate of SUSY particles at the LHC is very large unless the squarks
  and gluinos are very heavy. Looking for signals of direct production of gauginos
  and sleptons is difficult but not impossible (jet vetoes have to be
  used).
  It is very difficult to observe
  the heavier gauginos 
  unless they are strongly mixed because otherwise they are mainly
  Higgsino and therefore do not couple to the squarks and so cannot be
  produced in their decay.

  A future $\rm{e}^+\rm{e}^-$ collider would be a very powerful tool
  for precise measurements of masses and couplings provided that the
  energy is high enough. A few precise measurements made with such a
  machine could,
 in combination with the LHC measurements,  greatly constrain the
 underlying model.

  We are approaching  the end of an era. Low energy supersymmetry has
 been studied
  for the last twenty years, without any experimental verification.
  However the within the next eight years or so, the searches will
  reach high enough mass scales so that it will either be discovered
  or cease to be relevant to the problem of electro-weak symmetry breaking.

\section*{Acknowledgements}                            

We thank James Stirling and the other organizers of the Durham Workshop
on Beyond the Standard Model Physics where the talk on which this note
is based was given.

The work was supported in part by the Director, Office of Energy
Research, Office of High Energy Physics, Division of High Energy
Physics of the U.S. Department of Energy under Contract
DE--AC03--76SF00098.  Accordingly, the U.S.
Government retains a non-exclusive, royalty-free license to publish or
reproduce the published form of this contribution, or allow others to
do so, for U.S. Government purposes.

\end{document}